\documentclass[letterpaper, 10 pt, conference]{ieeeconf}
\IEEEoverridecommandlockouts
\usepackage{amsmath,amssymb,amsfonts}
\usepackage{textcomp}
\usepackage[utf8]{inputenc}
\usepackage[T1]{fontenc}
\usepackage[american]{babel}
\usepackage{graphicx}
\usepackage{scrhack} 
\usepackage{listings}
\usepackage{lstautogobble}
\usepackage{tikz}
\usepackage{tikzscale}
\usetikzlibrary{shapes, arrows, calc}
\usepackage{pgfplots}
\usepackage{pgfplotstable}
\usepgfplotslibrary{groupplots}
\usepackage{booktabs}
\usepackage{array}
\usepackage{numprint}
\usepackage{mathptmx} 
\usepackage{times} 

\usepackage{hhline}
\usepackage{tabularx}

\newcolumntype{C}{>{\centering\arraybackslash}X}

\newtheorem{rem}{Remark}

\usepackage[nolist,nohyperlinks]{acronym}
\begin{acronym}
	\acro{BO}[BO]{Bayesian optimization}
	\acro{GP}[GP]{Gaussian Process}
	\acro{KF}[KF]{Kalman Filter}
	\acro{MPC}[MPC]{Model Predictive Control}
	\acro{GD}[GD]{Gradient Descent}
	\acro{QP}[QP]{Quadratic Programming}
 	\acro{ML}[ML]{Maximum Likelihood}
\end{acronym}

\title{\LARGE \bf Automatic Parameter Tuning of Self-Driving Vehicles}

\author{Hung-Ju Wu, Vladislav Nenchev, and Christian Rathgeber
\thanks{Hung-Ju Wu, Vladislav Nenchev, and Christian Rathgeber are with BMW Group, Munich, Germany.
        {\tt\small hung-ju.wu@bmw.de, vladislav.nenchev@bmw.de, christian.rathgeber@bmw.de}}%
}

\begin{document}
\maketitle
\thispagestyle{empty}
\pagestyle{empty}

\begin{abstract}
Modern automated driving solutions utilize trajectory planning and control components with numerous parameters that need to be tuned for different driving situations and vehicle types to achieve optimal performance. This paper proposes a method to automatically tune such parameters to resemble expert demonstrations. We utilize a cost function which captures deviations of the closed-loop operation of the controller from the recorded desired driving behavior. Parameter tuning is then accomplished by using local optimization techniques. Three optimization alternatives are compared in a case study, where a trajectory planner is tuned for lane following in a real-world driving scenario. The results suggest that the proposed approach improves manually tuned initial parameters significantly even with respect to noisy demonstration data.
\end{abstract}


\section{Introduction} \label{section:introduction}

Many autonomous and automated driving systems are based on a modular
sense-plan-act architecture, where a model of the surroundings of the 
vehicle is created based on on- and off-board sensors, such that a 
motion trajectory for the vehicle can be planned, and executed by the actuators. Since the environment model is created based on noisy sensor measurements, solely relying on it to achieve human-like behavior in all driving scenarios can be suboptimal and imposes trade-offs during development. Therefore, upon implementing and deploying necessary algorithms on the target hardware, a desired dynamical behavior of an automated driving vehicle is typically achieved by varying parameters of its decision making components -- trajectory planners and controllers. A considerable manual effort is required to tune these parameters such that relevant specifications are met, a high performance is maintained for different vehicle platforms, and deployment time and cost throughout the whole product development life cycle is reduced.

We propose a method to automatically tune trajectory planning or controller parameters based on a limited number of recorded demonstrations from driving experts. For that, a cost function is defined, which captures the deviation of the recorded demonstrations from the closed-loop simulation of the to-be-tuned decision making components with a suitable system model. Then, three alternative optimization algorithms are suggested to obtain locally optimal parameters: \ac{GD}, an unscented \ac{KF}, and a \ac{ML} estimator. In a case study, we apply the method to tune the cost function weights of an optimal trajectory planner for lateral guidance of an automated driving vehicle based on resimulating field data.

The remainder of the paper is organized as follows: after discussing related work (Section\,\ref{sec:related_work}), in Section\,\ref{section:problemstatement} the parameter tuning problem is introduced. Section\,\ref{section:ParamOptiTrajPlanning} presents tuning approaches, which are then applied and discussed for a lateral trajectory planner in Section\,\ref{section:latTrajPlanning}. Finally, conclusions and possible future work are provided in Section\,\ref{section:conclusion}.

\section{Related work}\label{sec:related_work}

Parameter tuning is often addressed as a learning problem. Most learning-based approaches define a set of features such as speed, acceleration, jerk, distance to other vehicles, etc., included as linear combinations in a cost function, e.g., \cite{ziebart2008navigate}. However, some cost terms might be conflicting, e.g., small lateral position deviation from the center of the lane is achieved by a large jerk, but small jerks are favored in regard to driving comfort. Therefore, learning is often augmented by desired demonstrations, e.g., using techniques such as apprenticeship learning \cite{abbeel2008apprenticeship}, inverse optimal control \cite{2012-cioc}, or 
inverse reinforcement learning \cite{kuderer2015learning}. The effort in manual design of cost functions can be supported by inverse reinforcement learning or 
optimal control, where the cost function is learned with expert demonstrations \cite{friberg2021tuning, herman2015inverse, rosbach2019driving, shen2022inverse, sun2022inverse, Rudolf2022}. This 
provides the opportunity to even learn from failure demonstrations \cite{shiarlis2016inverse}. An alternative is to first train a deep neural network with reinforcement learning to determine the parameters of the cost function offline, and then adapt them online based on the offline learned policy \cite{zarrouki2021weights} for simulated curved trajectories. However, our goal to primarily use limited amounts of recorded field data to tune comfort driving behavior renders any of the aforementioned approaches intractable either due to high computational demands, the necessity of extensive data sets, or the availability of high-fidelity simulation models.

Parameter tuning can also be based on Bayesian optimization, where a mapping of the parameters to a pre-specified performance metric is learned, e.g., as a reward function using trial-and-error search \cite{Neumann2019}. Similarly, we also use a cost function to optimally tune the parameters, but we do not require episodical learning in a trial-and-error fashion.

Since the optimization-based tuning of controller parameters is typically non-convex, genetic algorithms can be used to estimate globally optimal parameters \cite{du2016development}. Particle swarm optimization shares the benefits of genetic algorithms, but usually provides better computational efficiency at the price of premature convergence. It was utilized for automatic parameter tuning for \ac{MPC} in \cite{kawai2007automatic}. It shows promising results in simulation for a simulated S-curve \cite{prakash2023evolutionary}, as well as with real data for car-following on an expressway \cite{lim2022automatic}. Differently from global optimization, we optimize the tuning cost function using local techniques to obtain generic (but not necessarily optimal) parameters.

Sampling-based estimators have been widely employed to tune parameters in the automotive domain. For example, Kalman filtering was effectively used to calibrate parameters of lane change controllers in simulation \cite{Menner2023}. However, Kalman filtering real data requires sufficient prior knowledge of the system, including the noise distribution and a good guess for the initial parameters, in addition to an accurate reference model. Some of these issues can be mitigated by limiting the considered driving scenario scope, e.g., by driving only at constant speed and for double lane changes on a test track \cite{wielitzka2018}. Similarly to these approaches, we use local improvements of the parameters with respect to a tuning cost function. In contrast to them, we consider state constraints and broader types of recorded field data.

\section{Problem Statement} \label{section:problemstatement}

Consider a discrete-time model of the vehicle behavior
\begin{equation}\label{eq:system}
\boldsymbol{x}_{t+1} = f(\boldsymbol{x}_t, \boldsymbol{u}_t)
\end{equation}
where $\boldsymbol{x}_t \in \{x\in\mathbb{R}^{n_x}: x_{min}\leq x\leq x_{max}\}$ is its state, $\boldsymbol{u}_t \in\{u\in\mathbb{R}^{n_u}: u_{min}\leq u\leq u_{max}\}$ is a control input, and $f$ is a general nonlinear function. Let a trajectory planner or controller with a set of the parameters $\boldsymbol{p}\in\{p\in\mathbb{R}^{n_p}: p_{min}\leq p\leq p_{max}\}$ be given, which takes the system state $\boldsymbol{x}_t$ at time step $t$ to provide an input $\boldsymbol{u}_t$ to the system (model). The control $\boldsymbol{u}_t$ can be resulting from solving an optimization problem, by querying a fixed mapping, multiplication of the state with a gain etc. For example, the parameters $\boldsymbol{p}$ can represent the gains of a PID controller, the weights of a cost function for \ac{MPC}, or weights and biases of a neural network. 

Let a demonstration of desired vehicle behavior $r$ be given as a trajectory $\boldsymbol{y}_{d, t}\in \mathbb{R}^{n_{y}}$ over time, i.e., $r:\boldsymbol{y}_d|_{[0,T]}$. Given a recorded demonstration $r$, the goal is to tune the parameters $\boldsymbol{p}$ of a trajectory planner or controller, such that it is capable of generating vehicle behavior with similar characteristics in other driving scenarios. In this context, error is defined as the deviation of a closed-loop simulation of the trajectory planner or controller with \eqref{eq:system} for a given parameter vector $\boldsymbol{p}$, yielding the trajectory $\boldsymbol{x}|_{[0,T]}$, from a corresponding recorded demonstration $r: \boldsymbol{y}_d|_{[0,T]}$. Assuming that for a recorded demonstration with duration $T$, this is captured by the positive term $\ell(\boldsymbol{y}_{d,t},\boldsymbol{x}_{t})$ and the cost function
\begin{equation}\label{eq:cost}
J(\boldsymbol{y}_d|_{[0,T]},\boldsymbol{x}|_{[0,T]})=\sum_{t=0}^{T} \ell (\boldsymbol{y}_{d,t},\boldsymbol{x}_{t}).
\end{equation}
The cost \eqref{eq:cost} is minimized to perform an automatic tuning as shown in the following.
\begin{rem}\label{sec:remark}
 Note that while the demonstration state $\boldsymbol{x}_{d, t}$ and the system state $\boldsymbol{x}_t$ may be defined differently, we assume that $\boldsymbol{x}_t$ is observable from $\boldsymbol{y}_{d, t}$. For example, $\boldsymbol{x}_t$ contains the curvature derivative for lateral vehicle motion in our case study (Section\,\ref{section:latTrajPlanning}), which is not directly measurable, but can be reconstructed using the considered model. Observability tests for various types of (non-)linear vehicle models can be readily performed \cite{Doumiati2013}. Further, to avoid that estimation is under-determined, we assume that the parameters $\boldsymbol{p}$ and the measured data $\boldsymbol{y}_d$ are chosen such that there exists a unique parameter that solves the optimal tuning problem.
\end{rem}

\section{Solution} \label{section:ParamOptiTrajPlanning}

An iterative optimization framework is proposed to obtain parameters that yield a closed-loop behavior of the trajectory planner or controller that resembles the recorded demonstration $r$. Figure \ref{fig:problem_formulation} depicts the scheme for one demonstration. Let the initial parameter vector be $\boldsymbol{p}_0$. For each iteration $k$ of the optimization, we compute the closed-loop trajectory $x|_{[0,T]}$. For that, at each time step $t\in[0, T]$, the model \eqref{eq:system} takes the control input from the trajectory planner or controller with parameters $\boldsymbol{p}_k$, and returns the next system state $\boldsymbol{x}_{t+1}$ (within the dashed rectangular box). Let $\boldsymbol{x}_{d,t} = C \boldsymbol{y}_{d,t}$, where the matrix $C$ is of appropriate dimension, and ensures observability properties as outlined in Remark~\ref{sec:remark}. Then, the cost \eqref{eq:cost} is obtained as the squared error between $\boldsymbol{x}|_{[0,T]}$ and $\boldsymbol{x}_d|_{[0,T]}$ weighted by the matrix $Q\in \mathbb{R}^{n_x\times n_x}$, i.e.,
\begin{align}\label{eq:detailed_cost}
 J_k (\boldsymbol{y}_d|_{[0,T]},\boldsymbol{x}|_{[0,T]})= \sum_{t=0}^{T} (\boldsymbol{x}_{t} - C\boldsymbol{y}_{d,t})^\top Q (\boldsymbol{x}_{t} - C\boldsymbol{y}_{d,t}).
\end{align}
In the following we assume that \eqref{eq:detailed_cost} is differentiable with respect to $\boldsymbol{p}$ to facilitate the use of gradient-based optimization methods, which are computationally efficient and typically produce good local estimates. 

Assuming that the specified variable bounds are not too restrictive or conflicting, for any bounded variable, we define an unbounded substitution variable, which is transformed back to the admissible range. For example, for the state $x_{1,t}\in[x_{1,min},x_{1,max}]$ and its substitution $\tilde{x}_{1,t}\in \mathbb{R}$, we use
\begin{equation} \label{eq:ukf_prediction}
\begin{split}
x_{1,t} = \frac{x_{1,max}\text{--}x_{1,min}}{2}\tanh{(\tilde{x}_{1,t})} + \frac{x_{1,min}+x_{1,max}}{2}.
\end{split}
\end{equation}
All state, parameter and input bounds are implemented analogously to \eqref{eq:ukf_prediction}. Therefore, the bounds do not need to be considered explicitly and unbounded optimization can be used. Using this transformation ensures that the solution lies within admissible bounds, improves problem feasibility, reduces the search space (potentially speeding up convergence) and allows incorporating practical limitations.

\tikzstyle{traj_plan} = [rectangle, 
minimum width=3cm, 
minimum height=1cm, 
text centered, 
text width=3cm, 
draw=black, 
fill=white]

\tikzstyle{plant} = [rectangle, 
minimum width=3cm, 
minimum height=1cm, 
text centered, 
text width=3cm, 
draw=black, 
fill=white]

\tikzstyle{plan_box} = [rectangle, 
minimum width=10cm, 
minimum height=3cm, 
text centered, 
text width=3cm, 
draw=black]

\tikzstyle{cost_calc} = [rectangle, 
minimum width=3cm, 
minimum height=1cm, 
text centered, 
text width=3cm, 
draw=black, 
fill=white]

\tikzstyle{param_opti} = [rectangle, 
minimum width=3cm, 
minimum height=1cm, 
text centered, 
text width=3cm, 
draw=black, 
fill=white]

\tikzstyle{sum} = [circle, 
node distance=1cm, 
draw=black]

\tikzstyle{arrow} = [thick,->,>=stealth]
\tikzstyle{pinstyle} = [pin edge={<-,thin,black}]

\begin{figure}[t!]
\centering
\vspace*{0.15cm}
\resizebox{0.49\textwidth}{!}{
    \begin{tikzpicture}[node distance=2cm]
    \node (traj_plan) [traj_plan] {Trajectory Planner/ Controller};
    \node (plant) [plant, right of=traj_plan, xshift=2cm] {Model};
    \node (cost_calc) [cost_calc, below of=plant, yshift=-0.5cm] {Cost Calculation};
    \node (param_opti) [param_opti, left of=cost_calc, xshift=-2cm] {Parameter Optimization};
    
    \draw [arrow] ([yshift=0.1cm]traj_plan.east) -- node[above] {u} ([yshift=0.1cm]plant.west);

    \draw [thick] (plant.south) - ++(0, -0.4);
    \draw [arrow] ($(plant.south)+(0,-0.4)$) -| (traj_plan.south);
    \draw [thick, dashed] ($(traj_plan.west)+(-0.25,1.22)$) rectangle ($(plant.east)+(0.25,-1.22)$);
    
    \draw [thick] ($(plant.east)+(0.25,0)$) - ++(0.5, 0) node[anchor=north, yshift=0.7cm] {$\boldsymbol{x}|_{[0,T]}$};
    \draw [arrow] ($(plant.east)+(0.75,0)$) |- (cost_calc.east);
    \draw [thick,<-] (cost_calc.south) -- ++(0, -0.5) node[right, yshift=0.2cm] {$\boldsymbol{y}_{d}|_{[0,T]}$};
    \draw [arrow] (cost_calc) -- node[above] {J} (param_opti);
    \draw [thick,-] (param_opti.west) - ++(-0.75, 0);
    \draw [arrow] ($(param_opti.west) + (-0.75,0)$) |- node[anchor=north, xshift=0.2cm, yshift=0.6cm] {$\boldsymbol{p}$} ($(traj_plan.west)+(-0.25,0)$);
    \end{tikzpicture}
}
\caption{Iterative parameter tuning approach with the simulated closed-loop trajectory $\boldsymbol{x}|_{[0,T]}$ of a trajectory planner or controller parameterized by $\boldsymbol{p}$, the input disturbance $\boldsymbol{z}_t$, the recorded trajectory of a demonstration $\boldsymbol{y}_{d}|_{[0,T]}$, and the tuning cost function $J$.}
\label{fig:problem_formulation}
\end{figure}
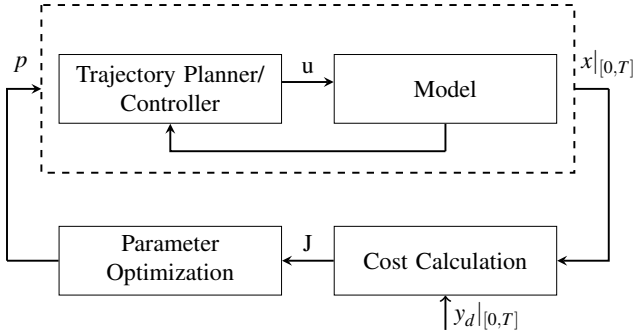
In the following, parameter optimization is addressed by three alternatives.

\subsection{Gradient Descent}\label{sec:grad}
Starting at the initial value $\boldsymbol{p}_0$, the gradient of the cost \eqref{eq:detailed_cost} is computed numerically, where the corresponding trajectory planner or controller with $\boldsymbol{p}$ is forward simulated with \eqref{eq:system} to obtain $\boldsymbol{x}|_{[0,T]}$. Then, with a sufficiently small learning rate $\gamma \in \mathbb{R}^+$ we move in the opposite direction of the gradient of the cost \eqref{eq:detailed_cost} and update the parameter at iteration step $k$ by
\begin{equation} \label{eq:gd}
\begin{split}
\boldsymbol{p}_{k+1} &= \boldsymbol{p}_k - \gamma \nabla_p J_k(\boldsymbol{x}_d|_{[0,T]},\boldsymbol{x}|_{[0,T]}).
\end{split}
\end{equation}
We assume that the updates of the parameters follow \eqref{eq:gd} without consideration of randomness, i.e., that the optimization is deterministic. A new parameter vector $\boldsymbol{p}_{k+1}$ with a corresponding cost $J_{k+1}$ is obtained, and the process is repeated until termination criteria are satisfied. The iterative optimization stops either when the maximal number of iterations $K$ is reached, or when the cost deviation between two consecutive iterations is below a specified threshold $\epsilon$, i.e., $|J_{k}-J_{k+1}|<\epsilon$. The learning rate can be adaptively updated in each iteration according to the range of each parameter to allow for a faster convergence.

\subsection{Unscented Kalman Filter}
The parameters $\boldsymbol{p}$ can alternatively be obtained by performing a simultaneous state and parameter estimation using a Kalman filter with the recorded demonstration data. For that, let the state of \eqref{eq:system} be augmented to $\boldsymbol{\tilde{x}}_{t}=[\boldsymbol{x}_t^\top,\boldsymbol{p}^\top]^\top$. Therefore, we use the matrix $\tilde{C} = [C^\top,0_{n_p\times n_y}^\top]^\top$, where $0_{n_p\times n_y}$ denotes a zero matrix of size $n_p\times n_y$ instead of $C$ in the cost\eqref{eq:detailed_cost}. Further, let $Q_e \in \mathbb{R}^{(n_x+n_p)\times (n_x+n_p)}$ be given by $Q_e = \text{blkdiag}(Q,\mu_e I_{n_p})$, where $I_{n_p}$ is an identity matrix of size $n_p$ and $\mu_e\in\mathbb{R}^+$ is a weighting parameter, such that the \ac{KF} approximates the cost \eqref{eq:detailed_cost} for the recorded demonstration. We assume constant parameters as a model for the process update, i.e., $\boldsymbol{p}_{n,t+1}=\boldsymbol{p}_{n,t}$, $n\in\{1,\ldots,n_p\}$. Then, using the dynamics \eqref{eq:system} for the original states $\boldsymbol{x}_t$ yields the augmented model for estimation
\begin{equation} \label{eq:ukf_sysmodel_y}
\begin{split}
\boldsymbol{\tilde{x}}_{t+1} &= \tilde{f}(\boldsymbol{\tilde{x}}_{t}, \boldsymbol{u}_{t})+\boldsymbol{w}_{t},\\
\boldsymbol{\tilde{y}}_{t} &= \tilde{C}\boldsymbol{\tilde{x}}_{t} + \boldsymbol{v}_{t},
\end{split}
\end{equation}
where the process noise $\boldsymbol{w}_t \propto \mathcal{N}(0, Q_e)$, and $\boldsymbol{v}_t \propto \mathcal{N}(0, R_e)$, $R_e \in \mathbb{R}^{n_y\times n_y}$ are assumed to be Gaussian. Then, since the function $\tilde{f}$ might be nonlinear, an unscented \ac{KF} is employed. The unscented \ac{KF} uses deterministic samples (called sigma points) around the mean, which are propagated and used to update the mean and covariance estimates of the parameters $\boldsymbol{p}$ and the states $\boldsymbol{x}_t$. Note that the \ac{KF} minimizes the (approximate) expected value of \eqref{eq:detailed_cost}. In this work, we utilize a standard unscented Kalman filter for estimation. We refer the interested reader to the detailed description of the prediction and update step presented in \cite{julier1997}. 

\subsection{Maximum Likelihood Estimation}
Another option to obtain $\boldsymbol{p}$ is to use \ac{ML} estimation. For that, the system model \eqref{eq:ukf_sysmodel_y} is used. As the overall cost \eqref{eq:detailed_cost} and all terms are positive or zero by definition, \eqref{eq:detailed_cost} is used as a likelihood function. Given the recorded demonstration $r$, we iteratively compute the parameters $\boldsymbol{p}_k$ by 
\begin{equation} \label{eq:mle_likelihoodeq}
\begin{split}
\hat{\boldsymbol{p}_k} &= \underset{\boldsymbol{p}_k}{\arg\max}(-\ln J_k(\boldsymbol{x}_d|_{[0,T]}, \boldsymbol{x}|_{[0,T]}|\boldsymbol{p}_k)).
\end{split}
\end{equation}
Similar to the previous section, let the initial augmented state estimate be $\boldsymbol{\hat{x}}_0\in\mathbb{R}^{n_x+n_p}$ and the initial covariance matrix estimate be $\boldsymbol{\hat{P}}_0 \in \mathbb{R}^{(n_x+n_p)\times (n_x+n_p)}$. By forward simulating the corresponding trajectory planner or controller with $\boldsymbol{p}_k$ and \eqref{eq:ukf_sysmodel_y} we obtain $\boldsymbol{x}|_{[0,T]}$, such that we can compute the cost $J$. By numerically estimating the gradient of the cost, the optimization problem \eqref{eq:mle_likelihoodeq} is solved using gradient descent similarly to Section~\ref{sec:grad}. Note that in contrast to Section~\ref{sec:grad}, ML estimation includes the underlying probabilistic assumptions about the data and we estimate both the parameters and the covariance. We refer the interested reader to \cite{Ljung1999} for further details on standard \ac{ML} formulations.

\section{Application to lateral trajectory planning} \label{section:latTrajPlanning}
The proposed methods are applied to a trajectory planner for lane keeping. It is used for the generation of trajectories for the lateral movement of a vehicle along a reference path. 

\subsection{Lateral Trajectory Planning}
\begin{figure}[b]
	\centering
    \includegraphics[width=0.48\textwidth]{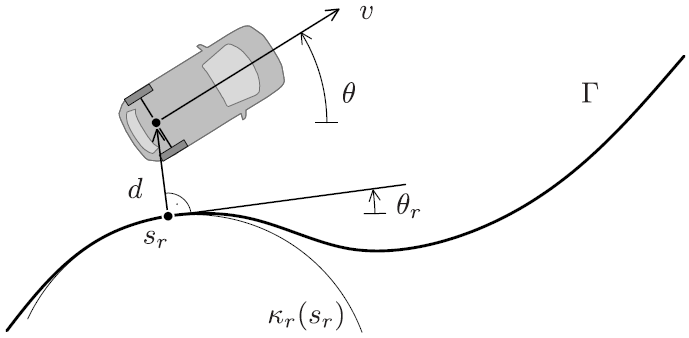}
	\caption{Kinematic vehicle model with respect to a given reference curve $\Gamma$ based on \cite{gutjahr2016lateral}.}
	\label{fig:kinematic_vehicle_model}
\end{figure}

The applied approach is based on \cite{gutjahr2016lateral}, where lateral trajectory planning is formulated as a linear time-varying constrained \ac{MPC} problem with a quadratic cost function similar to \cite{camacho2013model}. The lateral vehicle movement is described by linearized relative kinematics to a reference path (e.g. the center of the lane). As shown in Figure\,\ref{fig:kinematic_vehicle_model}, $d$ represents the signed normal distance between the reference curve $\Gamma$ and the rear axis center of the vehicle. $\theta$ is the vehicle orientation. 
Explicit consideration of lateral vehicle dynamics (e.g. by a dynamic single-track model) is omitted in favor an integrator chain to allow for an efficient real-time computation. The time derivative of the curvature change $\ddot \kappa$ is used as the system input $u$ to achieve a steering angle that can be continuously differentiated several times. This results in the system state $\boldsymbol{x} = [d, \theta, \kappa, \dot\kappa]^\top$. The reference curve is represented by its orientation $\theta_r$ and curvature $\kappa_r$ at the current base point, defined by the curve's arc length variable $s_r$.
The reference orientation serves as a disturbance input to the system $z=\theta_r$. $v$ describes the vehicle speed which is treated as a time-variant parameter. The system model for the lateral planner is transformed into discrete-time form with a sampling time $T_s$, leading to
\setlength{\arraycolsep}{5pt}
\begin{eqnarray} \label{system_discrete_lat}
\boldsymbol{x}_{t+1}&{}={}&
\begin{bmatrix}
1 & vT_s & \frac12v^2T_s^2 & \frac16v^2T_s^3\\
0 & 1 & vT_s & \frac12vT_s^2\\
0 & 0 & 1 & T_s\\
0 & 0 & 0 & 1
\end{bmatrix}\boldsymbol{x}_t\nonumber\\ 
&&{+}\:\begin{bmatrix}
\frac{1}{24}v^2T_s^4 \\
\frac16vT_s^3 \\
\frac12T_s^2 \\
T_s 
\end{bmatrix}u_t + \begin{bmatrix}
\text{--}vT_s\\
0\\
0\\
0
\end{bmatrix}{z}_t\nonumber,\\
\boldsymbol{y}_t &{}={}& \boldsymbol{x}_t.
\end{eqnarray}
The lateral planning objective is the weighted sum of the deviation of the states and the reference trajectory $\boldsymbol{x}_{r} = [0, \theta_{r,t}, \kappa_{r, t}, \dot\kappa_{r, t}]^\top $ and the inputs over the planning horizon $N$.
With $\boldsymbol{Q} = \mathrm{diag}(w_d, w_\theta, w_\kappa, w_{\dot\kappa})$ and $\boldsymbol{R} = w_u$, where $w_d$, $w_{\theta}$, $w_{\kappa}$, $w_{\dot \kappa}$ and $w_u$ are the weights of the system states and the control input, respectively, the overall MPC problem reads:
\begin{equation} \label{mpc_lat_obj}
\begin{aligned}
\text{min}_{u_{[0,N]}} &\sum_{n=0}^{N}([\boldsymbol{x}_n \text{--} \boldsymbol{x}_{r,n}]^\top \boldsymbol{Q}[\boldsymbol{x}_n \text{--} \boldsymbol{x}_{r,n}] + u_n^\top \boldsymbol{R}u_n),\\
\text{s.t. } & \forall n{\in}[0,N],\eqref{system_discrete_lat}, x_n{\in}[x_{min},x_{max}], u_n{\in}[u_{min},u_{max}].
\end{aligned}
\end{equation}
Since the cost is quadratic and the variables are linearly constrained, \eqref{mpc_lat_obj} can be solved efficiently by off-the-shelf \ac{QP} solvers.

\subsection{Implementation}
The recorded expert demonstrations are obtained by measuring the vehicle behavior when following the reference curve by a human driver. The system model \eqref{eq:system} is given by \eqref{system_discrete_lat}. The recordings contain motion trajectories with state $\boldsymbol{y}_{d} = [d_{d}, \theta_{d}, \kappa_{d}]^\top$, sampled with 80 ms. Since $\Dot\kappa$ is not measured, to obtain the desired state $\boldsymbol{x}_{d,t} = C \boldsymbol{y}_{d,t}$ we define
\begin{align*}
 C=\begin{bmatrix}
1 & 0 & 0\\
0 & 1 & 0\\
0 & 0 & 1\\
0 & 0 & 0
\end{bmatrix}.
\end{align*}
Clearly, the pair consisting of the system matrix of \eqref{system_discrete_lat} and $C$ is observable. The desired state $\boldsymbol{x}_{d}$ for this specific application is relative to the reference curve $\boldsymbol{x}_{r}$, which is recorded as a part of the demonstration. 

In this case study, we consider tuning the weights of the MPC problem, i.e., $\boldsymbol{p} = [w_d, w_\theta, w_\kappa, w_{\dot\kappa}, w_u]^\top$, which are particularly hard to tweak based on prior knowledge to resemble the demonstrated behavior. While by using our method it is possible to tune all parameters, certain parameters such as the planning horizon length and the sampling time have a significant impact on the feasibility and stability of gradient-based optimization and might require more complex handling. To optimize the parameters, the cost function \eqref{eq:cost} is given by
\begin{equation} \label{eq:opticostfunc}
\begin{split}
J &= \sum_{t=0}^{T} (\Delta d_t^2 + \Delta \theta_t^2 + \Delta \kappa_t^2 + \Delta\dot{\kappa}_t^2),
\end{split}
\end{equation}
where $\Delta d = d \text{--} d_{d}$, $\Delta\theta = \theta \text{--} \theta_{d}$, $\Delta\kappa = \kappa \text{--} \kappa_{d}$, and $\Delta\dot{\kappa} = \dot{\kappa} \text{--} 0$. All tuning algorithms are implemented in Python. The packages FilterPy and SciPy are used for the \ac{KF} and \ac{ML} approach, respectively. The \ac{MPC} problem is solved using the included \ac{QP} solver in CasADI \cite{Andersson2019}. At each time step, we solve the \ac{MPC} problem with the current reference path. The first value of the control vector is used to calculate the next state based on the vehicle model. This state is used as an initial state for solving the \ac{MPC} problem at the following time step, together with an updated reference path. The hyperparameters are chosen as follows:\\
$T_s = 0.3$\,s, $N=20$\\
            $\boldsymbol{x}_{min}=[-2, -1000, -0.04, -0.15]^\top$ \\
            $\boldsymbol{x}_{max}=[2, 1000, 0.04, 0.15]^\top$ \\
            $u_{min}=-0.07$, $u_{max}=0.07$ \\
            $\boldsymbol{p}_{min}$: $\numprint{e-6}[1,1,1,1,1]^\top$\\
            $\boldsymbol{p}_{max}$: $[\numprint{e5}, \numprint{e5}, \numprint{e5}, \numprint{e5}, \numprint{e-3}]^\top$\\
Maximal number of iterations $K= \numprint{e4}$\\
Learning rate $\alpha= 0.002$\\
Termination threshold $\epsilon= 0.001$\\
$Q$ = diag([1, 1, 1, $\numprint{e-8}$])\\
$\mu_e$= $\numprint{e-8} [1,1,1,1]^\top$\\
$R_e$= diag([0.0052, 0.0052, 0.0039, 0.0001])\\
$\boldsymbol{\hat{P}}_0$= diag([1, 1, 1, 1, 1, 1, 1, 1, 1])

\subsection{Results and discussion} \label{section:results}
\begin{figure}[tb!]
	\centering
    \vspace*{0.15cm}
    \includegraphics[width=0.48\textwidth]{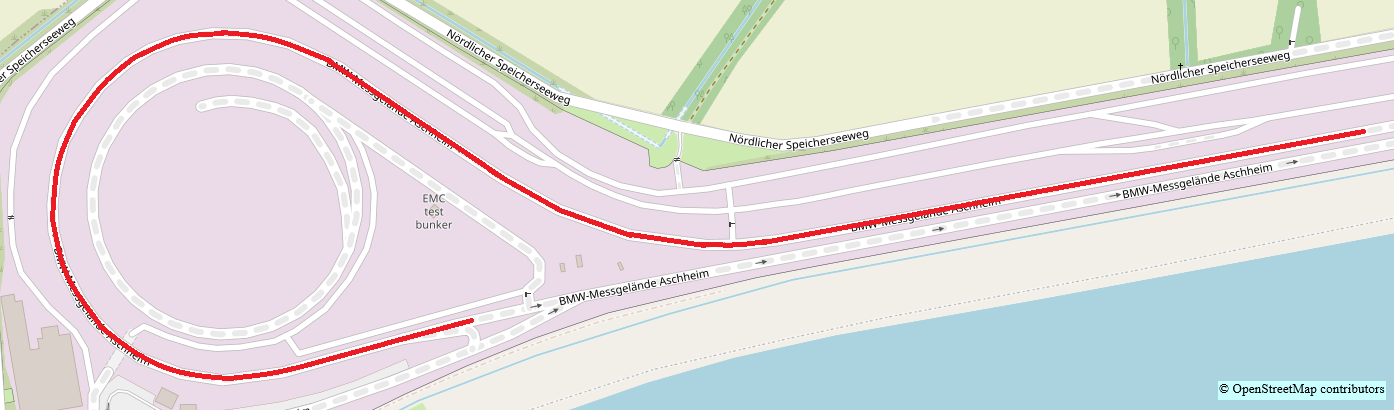}
	\caption{Bird's eye view of the relevant part of the BMW test track in Aschheim with the driving scenario path marked in red.}
	\label{fig:aschheim_m1}
\end{figure}

The proposed methods are applied for a driving scenario with varying road curvatures. The vehicle enters a right curve with around 90 km/h, then slows down to approximately 60 km/h and enters a tight left curve. The maneuver was carried out at the BMW test track (see bird's-eye view in Figure \ref{fig:aschheim_m1}).
The driver has driven manually in the center of the lane. The recording contains the movement of the vehicle relative to the reference path, in this case the center of the lane. This data is used as demonstrated behavior and as a look-ahead input to the previously described lateral trajectory planner. Note that parameter tuning is performed using replay of the recorded data. 

Table \ref{table:qp_weight} contains the initial parameter values $\boldsymbol{p}_0$, which have been tuned manually, as well as the values resulting from the three optimization methods (\ac{GD}, \ac{KF} and \ac{ML}).
\begin{table}[b]
\begin{center}
\caption{Initial and optimized values of $\boldsymbol{p}$.}
\label{table:qp_weight}
\begin{tabularx}{0.48\textwidth} { 
    | >{\centering\arraybackslash}X 
    | >{\centering\arraybackslash}X 
    | >{\centering\arraybackslash}X 
    | >{\centering\arraybackslash}X 
    | >{\centering\arraybackslash}X 
    | >{\centering\arraybackslash}X | }
    \hline
    & $w_d$ & $w_{\theta}$ & $w_{\kappa}$ & $w_{\dot\kappa}$ & $w_{u}$\\
    \hline \hline \hhline{} 
    $\boldsymbol{p}_0$ & 3.443 & 0.535 & 0.535 & 0.03 & $\numprint{2.26e-6}$ \\ 
    \hline
    $\boldsymbol{p}$:\ac{GD} & 9.59 & 0.536 & 0.535 & 0.03 & $\numprint{2.26e-6}$ \\
    \hline
    $\boldsymbol{p}$:\ac{KF} & 0.76 & 0.114 & 0.117 & 0.00719 & 0.0019 \\
    \hline
    $\boldsymbol{p}$:\ac{ML} & 3.93 & 1.64 & 0.01 & 0.01 & $\numprint{1e-7}$ \\ 
    \hline 
\end{tabularx}
\end{center}
\end{table}

\begin{figure*}[ht!]
\centering
\vspace*{0.15cm}
\input{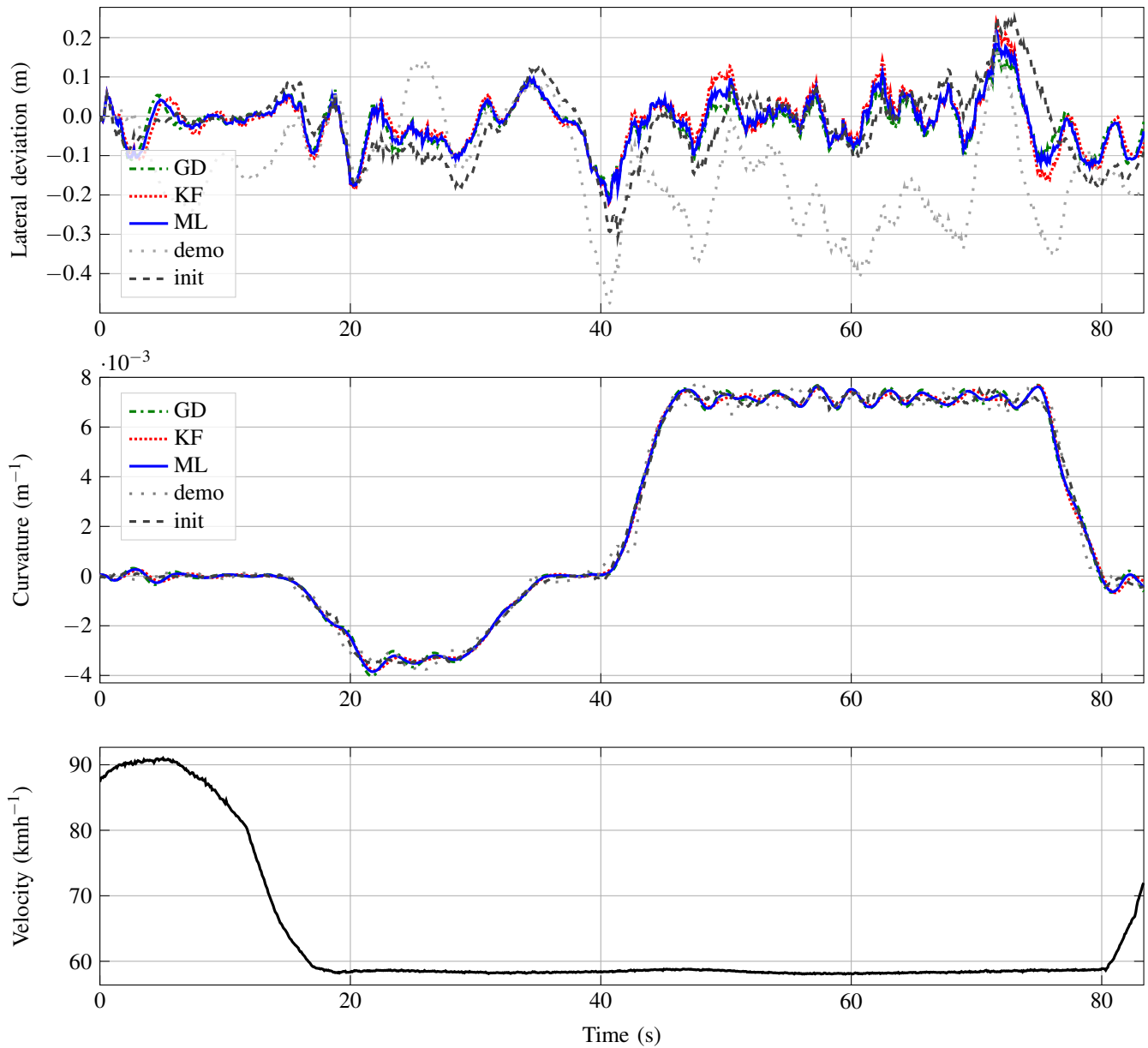}
\caption{Comparison of the resulting lateral deviation $d$ from the reference path and its curvature $\kappa$ using the tuned parameters with different optimization methods for the evaluation scenario.}
\label{fig:aschheim_m1_opti_comparison_d_kappa}
\end{figure*}

Figure \ref{fig:aschheim_m1_opti_comparison_d_kappa} shows the optimized behavior with $\boldsymbol{p}_0$ (denoted as "init") and $\boldsymbol{p}$ resulting from the three optimization methods. The recorded demonstration is denoted as "demo". Note that the demonstrated lateral deviation $d$ is not 0 as the driver slightly cuts the lane in favor of driving comfort. Furthermore, as only series vehicle sensors were used, the measurements are subject to partially substantial noise and measurement errors. In addition, the velocity is not constant throughout the driving demonstration. 

In the first half of the evaluation scenario (when driving on the straight segment and in a smaller right curve) all three methods show similar results in $d$. Note that the optimized behavior compromises well between the recorded demonstration and the reference path, e.g., in the time span 20-30\,s. In the second half of the trace, i.e., in the tight left curve, the optimized behavior exhibits significantly smaller lateral deviations compared to the demonstration, which contains lateral deviations of up to 45\,cm. This is caused by the \ac{MPC} aiming to minimize lateral deviations with respect to the reference path. As the addressed optimal tuning problem is not convex, despite starting the optimization with the same initial state, the three methods do not converge to the same local optimum.  Overall, \ac{ML} and \ac{GD} yield comparable behaviors, and \ac{KF} leads to slightly worse behavior. The values of state deviation integrals from the demonstration and the cost for different methods are summarized in Table \ref{table:compareaschheim_m1integral}. The demonstrated behavior is generally well-followed with the resulting parameters from each of the methods. This indicates the robustness and applicability of all three methods for noisy real data. 

\begin{table}[b!]
\begin{center}
\vspace*{0.15cm}
\caption{Integral of deviations from the recorded demonstration for different optimization methods in the evaluation scenario.}\label{table:compareaschheim_m1integral}
\begin{tabularx}{0.48\textwidth} { 
    | >{\centering\arraybackslash}X 
    | >{\centering\arraybackslash}X 
    | >{\centering\arraybackslash}X 
    | >{\centering\arraybackslash}X 
    | >{\centering\arraybackslash}X | }
    \hline
    & $\sum\Delta d_{d}^2$ & $\sum\Delta\theta_{d}^2$ & $\sum\Delta\kappa_{d}^2$ & Cost \\
    \hline \hline \hhline{} 
    init & 34.9 & 0.0289 & $\numprint{1.31e-4}$ & 34.9 \\
    \hline
    \ac{GD} & 29.6 & 0.0226 & $\numprint{1.61e-4}$ & 29.6 \\
    \hline
    \ac{KF} & 32.0 & 0.0179 & $\numprint{1.05e-4}$ & 32.0 \\
    \hline
    \ac{ML} & 28.2 & 0.0229 & $\numprint{1.71e-4}$ & 28.2 \\
    \hline
\end{tabularx}
\end{center}
\end{table}


\section{Conclusions} \label{section:conclusion}

We proposed a method to tune self-driving vehicle controller parameters automatically based on recorded demonstrations. For that, a cost function was introduced that captures the deviation of simulated runs of the controllers with a vehicle model from the desired behavior. The cost was efficiently minimized using any of the following local optimization methods: gradient descent, unscented Kalman filtering, or maximum likelihood estimation.  The proposed approach was applied to tune parameters of a lateral trajectory planner with respect to a real-world driving expert demonstration. All three methods were capable of optimizing the parameters of the trajectory planner with respect to the noisy recorded data at variable velocity. The maximum likelihood estimation method resulted in the smallest lateral deviations and overall cost. 

If differentiability assumptions for the cost do not hold, optimization can be done by using alternative techniques such as subgradient descent, proximal methods, or particle filters. Future work will focus on combining the proposed tuning approach with a global optimization framework. There is also potential for extending the methodology to encompass additional automated driving components. Another important research direction involves optimizing across various distinct driving scenarios concurrently.

\section*{Acknowledgement}
The authors would like to thank Mykyta Denysov for his help in preparing the data.

\bibliography{IEEEabrv,bibliography}

\end{document}